\documentclass[iop]{emulateapj}
\submitted{ApJ, accepted 09/12/2016}

\usepackage{graphics}
\usepackage{mathtools}
\usepackage{xspace}

\newcommand{\be}{\begin{equation}}
\newcommand{\ee}{\end{equation}}
\newcommand{\msun}{{\rm M}_{\odot}}
\newcommand{\kms}{{\rm km\,s^{-1}}}
\newcommand{\Mb}{{M}_{\rm b}}
\newcommand{\MHI}{{M}_{\rm HI}}
\newcommand{\Ms}{{M}_{\rm *}}
\newcommand{\Mh}{{M}_{\rm h}}
\newcommand{\jb}{{j}_{\rm b}}
\newcommand{\jHI}{{j}_{\rm HI}}
\newcommand{\jh}{{j}_{\rm h}}
\newcommand{\js}{{j}_{\rm *}}
\newcommand{\fj}{{f}_{\rm j}}

\newcommand{\fM}{{f}_{\rm M}}
\newcommand{\fg}{{f}_{\rm g}}
\newcommand{\Sigb}{{\Sigma}_{\rm b}}
\newcommand{\SigHI}{{\Sigma}_{\rm HI}}
\newcommand{\Sigs}{{\Sigma}_{\rm *}}

\newcommand{\Hi}{H{\sc\,i}\xspace}
\newcommand{\Vmax}{{V}_{\rm max}}
\newcommand{\VLOS}{{V}_{\rm LOS}}

\newcommand{\jcap}{J. Cosmol. Astropart. Phys.}

\begin{document}

\title{Angular Momentum of Dwarf Galaxies}

\author{Kirsty M. Butler$^1$}
\author{Danail Obreschkow$^1$}
\author{Se-Heon Oh$^{1,2}$}

\affiliation{$^1$International Centre for Radio Astronomy Research (ICRAR), M468, University of Western Australia, WA 6009, Australia\\
$^2$Korea Astronomy and Space Science Institute (KASI), Daejeon 305-348, Korea}

\begin{abstract}
We present measurements of baryonic mass $\Mb$ and specific angular momentum (sAM) $\jb$ in 14 rotating dwarf Irregular (dIrr) galaxies from the LITTLE THINGS sample. These measurements, based on 21cm kinematic data from the {\it Very Large Array} and stellar mass maps from the {\it Spitzer Space Telescope}, extend previous AM measurements by more than two orders of magnitude in $\Mb$. The dwarf galaxies show systematically higher $\jb$ values than expected from the $\jb\propto\Mb^{2/3}$ scaling of spiral galaxies, representative of a scale-free galaxy formation scenario. This offset can be explained by decreasing baryon mass fractions $\fM=\Mb/M_{\rm dyn}$ (where $M_{\rm dyn}$ is the dynamical mass) with decreasing $\Mb$ (for $\Mb<10^{11}\msun$). We find that the sAM of neutral atomic hydrogen (\Hi) alone is about 2.5 times higher than that of the stars. The $M$-$j$ relation of \Hi is significantly steeper than that of the stars, as a direct consequence of the systematic variation of the \Hi fraction with $\Mb$.
\end{abstract}

\maketitle


\section{Introduction}\label{section_introduction}

It has long been suggested that angular momentum (AM) and mass are the two most fundamental parameters in galaxy formation and evolution (\citealp{Fall1980,Quinn1988,Mo1998}), orchestrating other physical processes within galaxies that produce the properties we observe. In the Cold Dark Matter (CDM) dominated scenario of galaxy formation AM is acquired through tidal torques between neighbouring halos \citep{Peebles1969}. Approximate conservation of AM then dictates the ordered in-fall of gas into dark matter halos under dissipation of energy (\citealp{Fall1980}; \citealp{White1978}).

Empirically, the pioneering study by \cite{Fall1983} investigated the stellar mass $M_*$ and sAM $j_*$ of 44 spiral (Sb-Sc) galaxies and 44 elliptical galaxies -- a relation that was revisited by \cite{Romanowsky2012} with 67 spirals and 40 ellipticals spanning a larger range in morphology (E0-Sc). Total luminosities were used to derive $M_*$ whilst $j_*$ was approximated using kinematic tracers at specific optical radii. The sample formed two parallel trends of $j_*=qM_*^{\alpha}$ with exponents of $\alpha\approx2/3$ on the $M_*$-$j_*$ plane with elliptical galaxies containing three to four times less $j_*$ than spirals of equal $M_*$. 

High-resolution observations with integral field spectrographs (IFS) and/or radio interferometries allow us to obtain spectra for each pixel in spatially resolved images of galaxies on sub-kpc scales in the local universe. AM can then be integrated pixel-by-pixel. \cite{Obreschkow2014} (hereafter OG14) presented the first precision measurements of stellar and baryonic AM of 16 spiral (Sab-Scd) galaxies using neutral hydrogen (\Hi) kinematic maps in The \Hi Nearby Galaxy Survey (THINGS) (\citealp{Walter2008}), improving in accuracy of earlier measurements by an order of magnitude. The $\alpha\approx2/3$ trend found in \cite{Romanowsky2012} was reproduced by the sample. Galaxies with equal bulge mass fraction (defined as the mass fraction in excess of an exponential disk) however followed a $\alpha\approx1$ trend. This result refines the connection between sAM and Hubble morphologies, raising the question of how the $M$-$j$ relation behaves at lower masses where bulges are absent.

\begin{table*}
\begin{center}
	\label{tab}
	\begin{tabular}{ccccccccc}
		\hline\\[-2ex]
		Galaxy & D &$\MHI$ & $\Ms$ & $\Mb$ & $\jHI$ & $\mathrm{j_{\rm*}}$ & $\jb$ & $\Vmax$ \\
		 & [Mpc] & [$\mathrm{\log_{10}M_\odot}$] & [$\mathrm{\log_{10}M_\odot}$] & [$\mathrm{\log_{10}M_\odot}$] & [$\mathrm{\log_{10}kpc\ \kms}$] & [$\mathrm{\log_{10}kpc\ \kms}$] & [$\mathrm{\log_{10}kpc\ \kms}$] & [$\kms$]\\ [1ex]
		\hline
		\hline\\[-2ex]
		DDO50 & 3.40 & $9.25\pm_{8.28}^{7.70}$ & $8.34\pm_{7.35}^{6.78}$ & $9.42\pm_{8.42}^{7.86}$ & $2.48\pm_{1.42}^{1.62}$ & $1.88\pm_{0.85}^{0.93}$ & $2.45\pm_{1.41}^{1.61}$ & 32\\
		DDO52 & 10.30 & $8.87\pm_{8.42}^{8.08}$ & $8.02\pm_{7.34}^{7.23}$ & $9.05\pm_{8.56}^{8.25}$ & $2.71\pm_{2.09}^{2.42}$ & $2.06\pm_{1.14}^{1.16}$ & $2.68\pm_{2.08}^{2.42}$ & 66\\
		DDO70 & 1.30 & $8.67\pm_{7.42}^{6.16}$ & $7.48\pm_{6.07}^{4.98}$ & $8.82\pm_{7.56}^{6.31}$ & $2.17\pm_{1.12}^{1.13}$ & $1.55\pm_{0.35}^{0.08}$ & $2.15\pm_{1.11}^{1.12}$ & --\\
		DDO87 & 7.70 & $8.79\pm_{8.48}^{8.39}$ & $7.85\pm_{7.58}^{7.46}$ & $8.95\pm_{8.64}^{8.56}$ & $2.66\pm_{2.11}^{2.12}$ & $2.09\pm_{1.68}^{1.57}$ & $2.63\pm_{2.09}^{2.10}$ & 53\\
		DDO101 & 6.40 & $8.21\pm_{7.92}^{7.81}$ & $8.06\pm_{7.69}^{7.66}$ & $8.52\pm_{8.19}^{8.12}$ & $2.29\pm_{1.75}^{1.82}$ & $2.00\pm_{1.37}^{1.38}$ & $2.21\pm_{1.66}^{1.73}$ & 65\\
		DDO126 & 4.90 & $8.37\pm_{7.79}^{7.73}$ & $7.75\pm_{7.25}^{7.11}$ & $8.58\pm_{7.99}^{7.93}$ & $2.12\pm_{1.35}^{1.46}$ & $1.92\pm_{1.31}^{1.26}$ & $2.10\pm_{1.31}^{1.42}$ & 60\\
		DDO133 & 3.50 & $8.14\pm_{7.75}^{7.74}$ & $7.75\pm_{7.48}^{7.35}$ & $8.38\pm_{8.02}^{7.99}$ & $1.93\pm_{1.25}^{1.26}$ & $1.92\pm_{1.50}^{1.35}$ & $1.93\pm_{1.31}^{1.27}$ & 47\\
		DDO154 & 3.70 & $8.74\pm_{8.37}^{8.34}$ & $7.20\pm_{6.87}^{6.80}$ & $8.88\pm_{8.51}^{8.48}$ & $2.55\pm_{1.88}^{1.96}$ & $1.60\pm_{1.05}^{1.00}$ & $2.54\pm_{1.87}^{1.95}$ & 47\\
		DDO168 & 4.30 & $8.64\pm_{8.03}^{8.00}$ & $8.06\pm_{7.52}^{7.41}$ & $8.85\pm_{8.24}^{8.20}$ & $2.23\pm_{1.40}^{1.42}$ & $2.04\pm_{1.25}^{1.27}$ & $2.20\pm_{1.37}^{1.38}$ & 58\\
		DDO210 & 0.90 & $6.80\pm_{5.96}^{5.74}$ & $6.18\pm_{5.55}^{5.12}$ & $7.00\pm_{6.14}^{5.95}$ & $1.12\pm_{0.56}^{1.00}$ & $0.75\pm_{0.17}^{0.40}$ & $1.08\pm_{0.51}^{0.95}$ & 17\\
		DDO216 & 1.10 & $7.09\pm_{6.26}^{6.05}$ & $7.20\pm_{6.47}^{6.16}$ & $7.51\pm_{6.68}^{6.47}$ & $1.31\pm_{0.50}^{0.64}$ & $1.17\pm_{0.29}^{0.42}$ & $1.25\pm_{0.37}^{0.51}$ & 17\\
		NGC2366 & 3.40 & $8.99\pm_{7.56}^{7.15}$ & $8.30\pm_{7.06}^{6.46}$ & $9.18\pm_{7.74}^{7.34}$ & $2.44\pm_{1.06}^{1.16}$ & $2.12\pm_{0.93}^{0.63}$ & $2.41\pm_{1.02}^{1.12}$ & 59\\
		UGC8508 & 2.60 & $8.00\pm_{7.11}^{6.39}$ & $6.97\pm_{6.07}^{5.36}$ & $8.16\pm_{7.23}^{6.55}$ & $2.67\pm_{1.81}^{2.07}$ & $1.59\pm_{0.63}^{0.85}$ & $2.64\pm_{1.80}^{2.05}$ & 128\\
		WLM & 1.00 & $8.19\pm_{7.17}^{6.80}$ & $7.51\pm_{6.52}^{6.11}$ & $8.39\pm_{7.34}^{6.99}$ & $2.21\pm_{1.17}^{1.32}$ & $1.85\pm_{0.80}^{0.42}$ & $2.17\pm_{1.14}^{1.29}$ & 39\\
		\hline
	\end{tabular}
	\caption{Measured values for the 14 dIrr galaxies in this letter. Distances are taken from \cite{Oh2011} whilst masses, sAM and $V_{\rm max}$ are measured as in section 2.2 with the contribution from He only included in the baryon values. Upper and lower uncertainties on mass and sAM are found by combining distance errors with the 16\% and 84\% quantiles respectively, calculated from a 1000 iteration jack-knife resampling.}
\end{center}
\end{table*}

Both theoretically and observationally the low mass ($\Mb<10^{10}\msun$) end of the $M$-$j$ plane is not well understood. At these masses most star-forming galaxies are \Hi dominated (\citealp{Maddox2015}). Stars and molecular gas are subdominant and centrally concentrated, meaning that AM locked up at large radii is more reliably measured in \Hi kinematics. This work investigates the $M$-$j$ momentum relation of 14 dwarf irregular (dIrr) galaxies taken from the Local Irregulars That Trace Luminosity Extremes, The \Hi Nearby Galaxy Survey (LITTLE THINGS: \citealp{Hunter2012}). In section 2 we will lay out the sample of dwarf galaxies and measurement of their AM and mass. In section 3 we present a discussion on the mean $\Mb$-$\jb$ plane, scatter of this relation and the \Hi, stellar and baryon component $M$-$j$ relations. Conclusions are made in section 4.

\section{Observations of Mass and AM}
\subsection{The Sample and Data}
To investigate AM at the low mass end we have taken 14 local dIrr galaxies from the LITTLE THINGS sample. We exclude the four blue compact dwarfs present in LITTLE THINGS as well as galaxies lacking 3.6$\mu m$ {\it Spitzer} data and galaxies with inclinations less than $40^\circ$. Furthermore, the extremely irregular velocity map of DDO155 and poorly resolved velocity profiles of NGC4163, NGC1569 and CVnIdwA also exclude them from our analysis.

We make use of high spectral ($\leq2.6\kms$) and angular (${\sim}6''$) resolved \Hi kinematic data (two spatial dimensions and one velocity dimension) obtained with the {\it National Radio Astronomy Observatory} (NRAO) {\it Very Large Array} (VLA) and $3.6\mu m$ mid-infrared {\it Spitzer}/{\it IRAC} images. Together these data sets probe the majority of baryons within our galaxies \citep{Maddox2015} and allow us to make reliable measurements of the sAM ($j\equiv J/M$) in the stellar and \Hi discs. Total baryon mass is calculated by $\Mb=\Ms+1.36\MHI$ where the 1.36 factor accounts for the helium fraction at $z=0$ \citep{deBlok2008}. We neglect molecular hydrogen (H2), since CO measurements of nearby dwarf irregulars, whether identified by morphology \citep{Obreschkow2009a} or stellar mass ($M_*<10^9\msun$, \cite{Boselli2014}), suggest $M_{\rm H2}/\MHI<0.1$, even if accounting for low metallicity in the CO-H2 conversion. These low molecular gas fractions can be explained by the relatively low surface densities of  (non-compact) dwarfs \citep{Obreschkow2009b}, backed-up by inefficient H2-formation at low metallicities \citep{Lagos2011, Xie2016}.

The full sample along with their measured values can be found in Table \ref{tab}. 

\subsection{Extracting Radial Profiles}

\begin{figure*}
	\includegraphics[width=\textwidth]{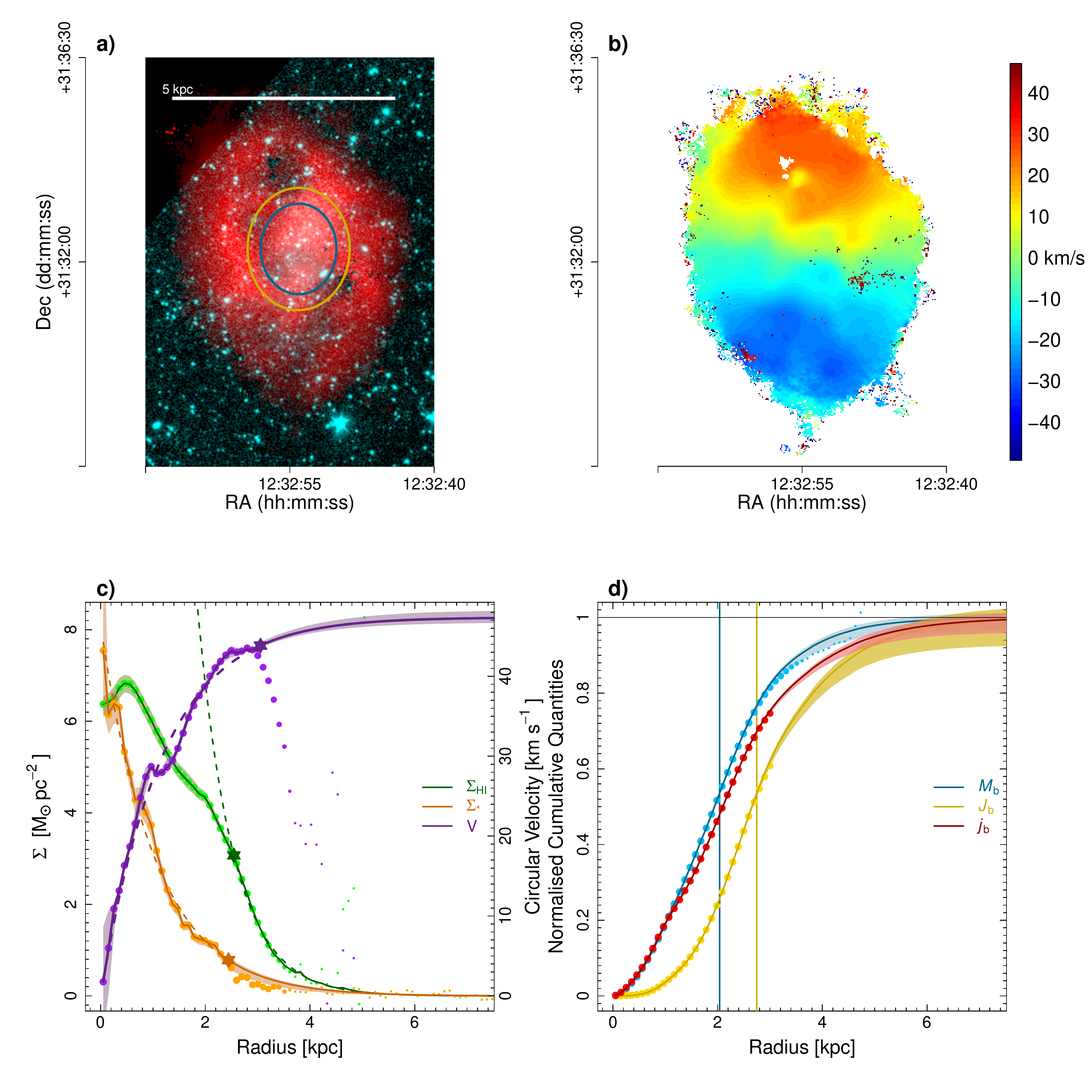}
	\caption{\textbf{(DDO133)} \textbf{a)} Composite raw \Hi-stellar intensity map with blue ellipse marking half mass radius and yellow ellipse marking half AM radius. \textbf{b)} Raw line of sight velocity map. \textbf{c)} Radial profiles of \Hi (green), stellar (orange) and circular velocity (purple) for the raw (dots, scaled in size by fraction of filled pixels), fitted (dashed) and hybrid (solid) profiles. Orange star lies at $10\%$ the maximum stellar density whilst placement of the green star is chosen by eye where raw data past this radius is used in the \Hi fit. Similarly the purple star is chosen by eye where raw data past this radius is not used in the velocity fit. Shaded regions show the $16\%$ and $84\%$ quantiles from a 1000 iteration jack-knife resampling. \textbf{d)} Cumulative radial profiles of baryon mass (blue), AM (gold) and sAM (red) for the raw (dots, scaled in size by fraction of filled pixels) and hybrid (solid) data. Shaded regions show the $16\%$ and $84\%$ quantiles from a 1000 iteration jack-knife resampling \citep{Quenouille1949}. Blue and yellow vertical lines indicate the half mass and AM radii respectively. }
	\label{DDO133}
\end{figure*}

As explained in this section, our method to measure the mass and AM uses a model of flat, axially symmetric disks. Measurement uncertainties (given in Table \ref{tab} and Fig.\ref*{fMmodels}) are estimated via a jack-knifing method: we only fit the model to a random half of the density and velocity data, and repeat this fit 1000 times with a different random seed at each iteration. The uncertainties are then estimated from the model deviations across all iterations \citep[following][]{Quenouille1949}. These uncertainties are accurate if the galaxies satisfy the model assumptions, but care must be taken in the case of model deviations such as irregular structures, warping and flaring. Since such density and velocity asymmetries will be picked up differently at each jack-knife iteration, they will automatically yield larger uncertainties. However, we caution that these uncertainties might only be lower limits, since most irregularities are spatially correlated across many pixels.
	
	Our five-step method optimally combines measurements with inter- and extrapolations, where data is missing.

\textit{Step 1 -- Data maps:}
Each \Hi data cube was run through a robust fitting process whereby the spectrum of each spaxel undergoes a simple parabolic background subtraction and then fitted with a Gaussian. The fitted intensity (S), line of sight (LOS) velocity ($\VLOS$) and velocity dispersion ($\sigma$) of each spaxel form 2D maps in the RA-Dec plane. A composite \Hi and stellar intensity map of DDO133 is shown in Fig.\ref{DDO133}a and the corresponding \Hi velocity map in Fig.\ref{DDO133}b.

Pixels are automatically rejected if S is less than the background root mean square (RMS) value measured for the entire galaxy \citep{Hunter2012}, if $|\VLOS|>500 \kms$, if $\sigma$ is less than two velocity bins or larger than the full velocity band-width or if the signal to noise ratio of the pixel is less than three. The intensity map is then converted from $\rm Jy\ beam^{-1}$ to $\rm M_{\odot}\ pixel^{-1} $ and the 3.6$\rm\mu m$ stellar images are converted from ${\rm MJy\ sr^{-1}}$ to ${\rm M_{\odot}}$ using a constant mass to light ratio of $1\rm M_{\odot}\ L_{\odot}^{-1}$ (for consistency with OG14). A bilinear interpolation is used to re-grid the stellar images to match the pixel size and dimension of the \Hi maps. 

Position angles, system velocities and inclinations taken from \cite{Oh2015} are used to compute the deprojected radii $r$ and circular velocity $V$ in each pixel. This involves simple trigonometry as explained by OG14 in Appendix B of their paper. Pixels of position angle within $10^{\circ}$ to the minor axis are dominated by radial motion due to turbulence and contain little information of the rotation. Pixels in this region are removed from the analysis.

\textit{Step 2 -- Data profiles:}
We bin pixels in the \Hi surface density $\Sigma_{\rm HI}$, stellar density $\Sigma_*$ and velocity maps into concentric ellipses of constant deprojected radius. Radial density profiles are extracted by taking the mean of $\Sigma_{\rm HI}$ in each ellipse and the median of $\Sigma_{\rm*}$ in each stellar ellipse to filter out the effects of foreground stars. The median V in each ellipse is used as this provides a smoother and more physical velocity profile than the mean.
	
\textit{Step 3 -- Model profiles:}
Assuming the background is fully subtracted in step 1 the \Hi density profiles are fitted such that $\SigHI(r)=\Sigma_{\rm HI,0}(2\pi r_{\rm HI}^2)^{-1}e^{-r/r_{\rm HI}}$. Many of the \Hi density profiles exhibit plateaus or dips at small radii where \Hi gas has been converted to molecular gas or ionised due to stellar feedback. Therefore, a radius is chosen by eye (green star in Fig.\ref{DDO133}c) for which points inside that radius are not included in the fitting process. Next we fit $\Sigs(r)=\Sigma_{\rm*,0}(2\pi r_{\rm *}^2)^{-1}e^{-r/r_{\rm *}}+\Sigma_{\rm bg}$ to the stellar density profile, allowing some vertical offset $\Sigma_{\rm bg}$ to be subtracted as background light. Due to the low signal to noise in the outer stellar disk we visually select a radius (orange star in Fig.\ref{DDO133}c) for which data beyond this radius is not used in the fitting process. The velocity profile is fitted via $V(r)=\Vmax(1-e^{-r/r_{\rm flat}})$, allowing us to extract $\Vmax$. Again a radius is chosen by eye (purple star in Fig.\ref{DDO133}c) for which points beyond this radius are not used in the fitting process. Many galaxies exhibit warping in their disks (see \citealp{Oh2015}) and therefore non-constant inclinations. This causes the velocity profile to artificially curve or wiggle as in Fig.\ref{DDO133}c forcing us to use only the inner velocity curve in our fit. This has little effect on the total sAM measurements since at this radius $j(r)$ is nearly converged (see red curve in Fig.\ref{DDO133}d). The model profiles are shown in Fig.\ref{DDO133}c as dashed lines.

\textit{Step 4 -- Hybrid Profiles:}
Hybrid maps are formed by combining raw data with the fitted profiles. Empty \Hi density map pixels lying within the green star's radius are replaced with the mean \Hi density in their ellipse. At larger radii the empty pixels are replaced with the fitted values corresponding to their exact r, allowing a smooth transition from data to model. All empty pixels in the velocity map are treated the same, replaced by a fitted value calculated for their exact {\it r}. As in step 3 we measure radial profiles, using the mean \Hi density and median circular velocity in each concentric ellipse. The hybrid stellar radial profile contains raw values at radii less than the orange star and model values at larger radii. Hybrid profiles are extended out to 15 times the scale radii $r_{\rm HI}$, predicting mass and AM out to larger radii than the extent of observational data (solid lines in Fig.\ref{DDO133}c).  

\textit{Step 5 -- Final values:}
Raw and hybrid total baryon density profiles are simply $\Sigb=\Sigs+1.36\SigHI$ where the 1.36 factor accounts for the He fraction ($M_{\rm He}/\MHI=0.36$) at $z=0$. Density radial profiles are converted to mass radial profiles $\Delta M(r)$ to calculate cumulative mass, AM  and sAM profiles using,
\begin{subequations}
	\begin{align}
	M(r)=&\underset{r_i<r}{\Sigma}\Delta M_i\\
	J(r)=&\underset{r_i<r}{\Sigma}\Delta M_iv_ir_i\\
	j(r)=&J(r)/M(r)
	\end{align}
\end{subequations}
where the subscript i denotes the ellipse of radius $r_i$ in the galaxy plane. Total integrated values are given by $M\equiv M(\infty)$, $J\equiv J(\infty)$ and $j\equiv j(\infty)$. Raw mass and AM cumulative profiles are corrected for the fraction of empty pixels in each ellipse. Hybrid values measured for the gas, stellar and baryonic $M$, $J$, and $j$ can be found in Table \ref{tab}. The $\Mb(r)$, $J_{\rm b}(r)$ and $\jb(r)$ profiles are shown in Fig.\ref{DDO133}d.

\section{Results: $M$-$j$ Plane}
\begin{figure}
	\vspace{-7mm}\includegraphics[width=\columnwidth]{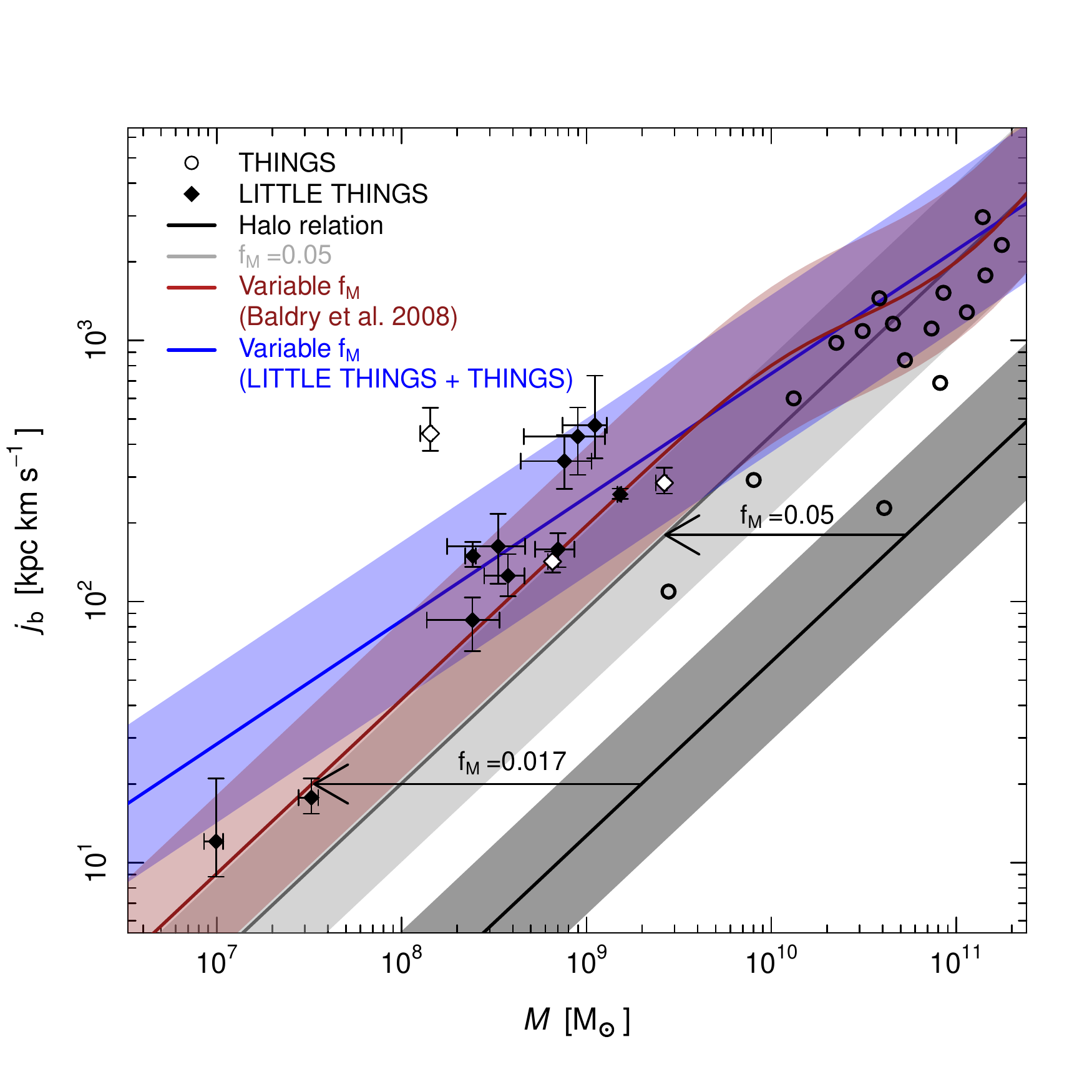}\vspace{-2mm}
	\caption{$\Mb$-$\jb$ relation of the 14 dIrr galaxies from this work (diamonds) and 16 spiral galaxies from OG14 (circles) compared to models for disks formed in spherical CDM halos with differing $\fM$ values. A constant $\fM$ estimated for Milky Way mass galaxies predicts the light grey region but only agrees with the THINGS galaxies. A decreasing $\fM$ with decreasing $\Mb$ better agrees with the deviated dwarf relation as this shifts galaxies left on the $M$-$j$ plane. The three White diamonds correspond to the LITTLE THINGS galaxies not included in the blue $\fM(\Mb)$ fit.}
	\label{fMmodels}
\end{figure}

Fig.\ref{fMmodels} displays the $\Mb$-$\jb$ plane with the 14 dIrr galaxies from this work (diamonds) and 16 spiral galaxies from OG14 (dots). The relation for dwarfs displays a similar scatter, but systematic offset to the relation for spirals (\citealp{Fall1980,Romanowsky2012}; OG14). We shall now explain these findings in a basic, CDM dominated galaxy formation framework \citep{White1978}. 

\subsection{Mean Baryonic $M$-$j$ relation} 

Spherical CDM halos exhibit the first order relationship $\jh\propto\lambda \Mh^{3/2}$ \citep{Mo1998} between mass $\Mh$, sAM $\jh$ and the spin parameter $\lambda$ \citep{Peebles1969}. The mass of baryons $\Mb$ that form the disk makes up a fraction $\fM\equiv\Mb/\Mh<1$ of the total mass. Assuming that the sAM fraction $\fj\equiv \jb/\jh$ is unity, as shown by modern simulation (within 50\%) \citep[e.g.][]{Stewart2013} and observations \citep{Fall1980}, it follows (OG14),

\begin{equation}
\frac{\jb}{10^3\mathrm{kpc\ km\ s^{-1}}}=1.96\lambda \fM^{-2/3}\Big[\frac{\Mb}{10^{10}\mathrm{M_\odot}}\Big]^{2/3},
\label{jb}
\end{equation}
where scatter about this relation is accounted for in the halo spin parameter $\lambda$ \citep{Steinmetz1995}. Fig.\ref{fMmodels} (grey line) is the $\Mb$-$\jb$ relation of Eq.~(\ref{jb}), for a constant $\lambda\approx0.03$ (typical of CDM halos; \citealp{Bullock2001}) and $\fM\approx0.05$ (typical for local Milky Way mass disks). The grey band represents 80\% of the skewed distribution of $\lambda$ \citep{Bullock2001} and agrees well with the THINGS galaxies. The LITTLE THINGS sample fall systematically above this relation, suggesting $\lambda$, $\fj$ or $\fM$ must vary with $\Mb$. However, cosmological simulations suggest $\lambda$ is about independent of $\Mb$ \citep{Knebe2008} and $\fj\approx1$ over any halo evolution deprived of major mergers \citep{Stewart2013}.

Assuming a universal baryon fraction of 17\% (relative to baryons and CDM), we can rewrite $\fM=0.17\epsilon$ where $\epsilon$ is the efficiency of the halo to form a baryonic disk. It is well established that $\epsilon$ (and therefore $\fM$) peaks for galaxies of Milky Way size ($\approx10^{11}\mathrm{M_\odot}$) (\citealp{Baldry2008}, \citealp{Behroozi2013}) where the gravitational potential wells are deep enough to retain baryons heated and accelerated by stellar feedback (Supernovae, stellar winds etc.). With decreasing baryon mass $\fM$ decreases to about $\fM\approx0.017$ at $\Mb=10^9M_\odot$. Applying this varying $\fM(\Mb)$, as given by \cite{Baldry2008}, the $\Mb$-$\jb$ relation becomes the red line and shading in Fig.\ref{fMmodels}. While this relation shows a better agreement with our dwarf galaxies, it still falls slightly below most measurements. This is likely due to the fact that \cite{Baldry2008} only measured $\fM$ down to galaxy masses around $\Mb=10^9M_\odot$. Therefore, we also estimate $\fM=\Mb/\Mh$ directly from the LITTLE THINGS and THINGS data using dynamical masses calculated via equation 3 derived by \cite{Oh2011},

\begin{equation}
M_{\rm dyn}/{\rm M_\odot}\simeq3.29\cdot10^5 [V_{\rm 200}/\kms]^3
\label{M200}
\end{equation}
where $V_{\rm200}$ is assumed approximately equal to the velocity at the largest measured radii; for LITTLE THINGS this is simply $V_{\rm max}$. UGC8508 is left out of this analysis as it is an outlier in the $M$-$j$ relation, along with DDO50 and DDO70 which have uncertain $V_{\rm max}$ values due to strong disk warping. We apply a log-log fit to $\fM(\Mb)$ and find $\fM(10^{11}M_\odot)=0.043$, decreasing to $\fM(10^9M_\odot)=0.003$, lower than that found by \cite{Baldry2008}. Our fitted $\fM$ values predict the region in blue and improves in the $10^8M_\odot$-$10^9M_\odot$ range.

\subsection{Scatter around the baryonic $M$-$j$ relation}
OG14 found the scatter about the spiral $M$-$j$ relation strongly correlated with bulge mass fraction $\beta$ with galaxies of equal $\beta$ following a $j\propto M$ trend. The full sample of dwarf galaxies in this work lay above the $\beta=0$ trend (no bulge), spinning to fast and sitting in gravitational wells too shallow for bulges to form in situ. As a replacement morphological tracer we computed the asymmetry {\it A} in the intensity ({\it S}) and velocity ({\it V}) maps of our sample, using a simple algorithm by \cite{Schade1995}. We first cropped the maps out to a radius where the mean HI density dropped below $0.5\msun\ \mathrm{pc^{-2}}$ then rotated the maps by $180^\circ$ ($S^{180}$ and $V^{180}$) to measure $A_{\rm S}=(\Sigma|S_{ij}-S^{\rm 180}_{ij}|)(2|S_{ij}|)^{-1}$ and $A_{\rm V}=(\Sigma|V_{ij}+V^{\rm 180}_{ij}|)(2|V_{ij}|)^{-1}$, where $i$ and $j$ are the pixel positions. There was no obvious trend in our data and further searches for correlation with other galaxy parameters such as gas fraction, specific star formation rate or depletion time showed no statistical significance. This however, is most probably due to the scatter of our data being dominated by uncertainty. 

\subsection{Component $M$-$j$ relations}
\begin{figure}
	\includegraphics[width=\columnwidth]{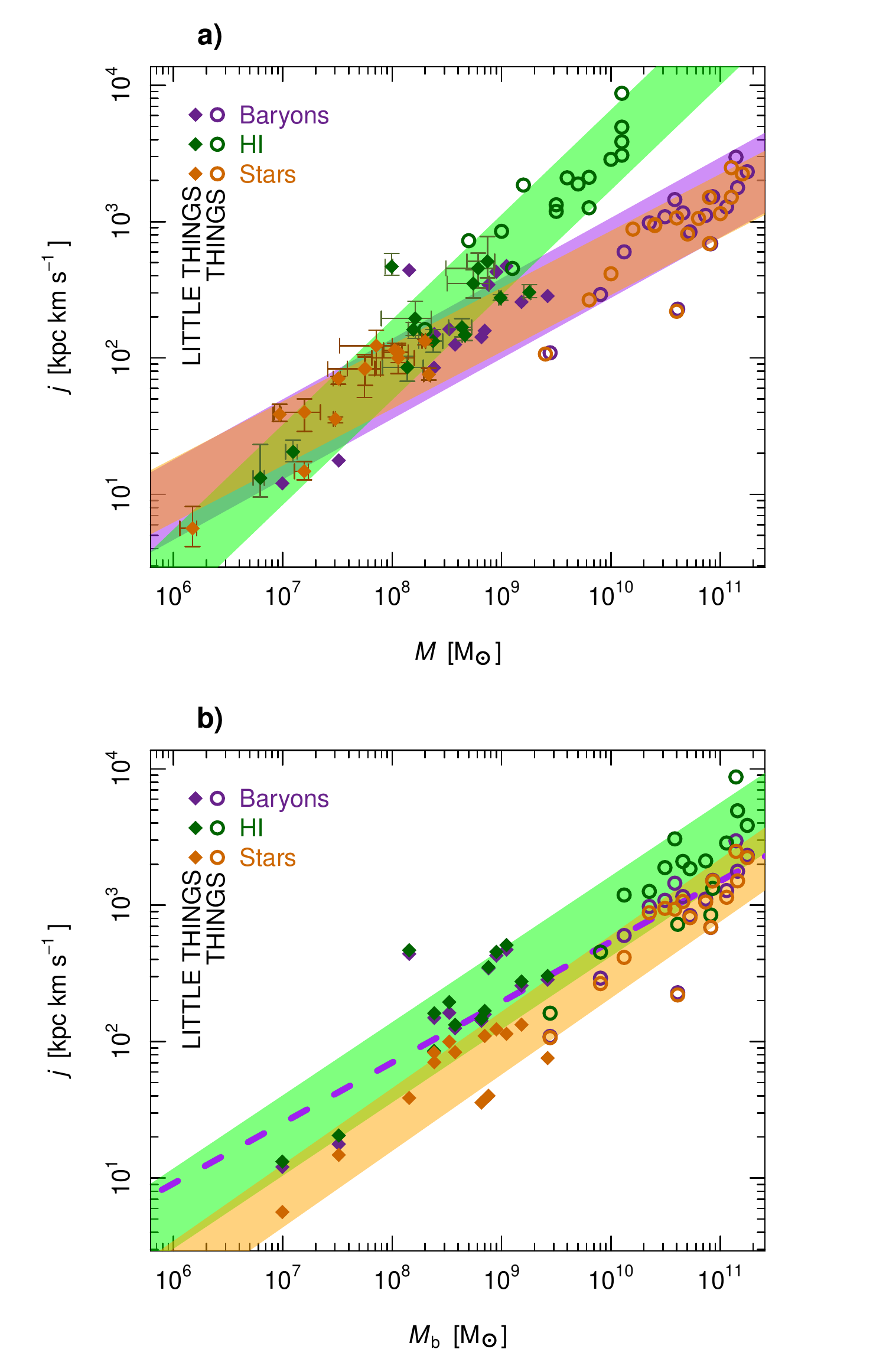}
	\caption{\textbf{a)} \Hi (green), stellar (orange) and baryon component $M$-$j$ relations for the THINGS (circles) and LITTLE THINGS (diamonds) galaxies with shaded regions (coloured respectively) outlining the 1$\sigma$ scatter from each component fit. \textbf{b)} \Hi (green), stellar (orange) and baryon component sAM as a function of $\Mb$ for the THINGS (circles) and LITTLE THINGS (diamonds) galaxies with shaded regions (coloured respectively) outlining the 1$\sigma$ scatter from each component fit.}
	\label{Mj}
\end{figure}

Fig.\ref{Mj}a compares the \Hi (green), stellar (orange) and total baryon (purple) $M$-$j$ relations with their respective one standard deviation scatters. \Hi mass $\MHI$ increases significantly faster with \Hi sAM $\jHI$ than that of stellar mass $\Ms$ with stellar sAM $\js$ and $\Mb$ with $\jb$. Despite the appearance of these relations, $\jHI>\jb>\js$ is always true. When plotted as a function of $\Mb$ (as in Fig.\ref{Mj}b), the \Hi and stellar trends lie almost parallel with \Hi components on average holding 2.5 times more sAM. Stars preferentially form in the central regions of the galactic disk, where low AM \Hi has sufficiently cooled and collapsed to form molecular clouds, setting up the stellar disks with systematically lower AM. Further low AM \Hi is lost through stellar feedback heating up and removing matter from the disk and new high AM \Hi accreted at large radii over time leads to the overall $\jHI>\jb>\js$ pattern we observe. At low masses galaxies have high gas fractions $\fg$ (\citealp{Maddox2015}) and the \Hi components therefore dominate the $\Mb$-$\jb$ trend in this regime. At higher masses where star formation is more efficient the stellar component dominates, resulting in a shallower trend for baryons. Thus, it is the variation of $\MHI$/$\Ms$ with $\Mb$ that causes the $\MHI$-$\jHI$ relation to be steeper than the $\Ms$-$\js$ relation.

\section{Conclusion}

We have presented measurements of mass and sAM for the \Hi, stellar and baryon components of 14 dIrr galaxies from the LITTLE THINGS sample. High resolution \Hi kinematic data and $3.6\mu$m {\it Spitzer} maps are combined with kinematic models to form hybrid maps allowing us to accurately integrate the full mass and AM. These measurements extend on previous AM measurements by more than two orders of magnitude in $\Mb$ ranging $10^6-10^9\mathrm{M_\odot}$. The sample is found to deviate from the spiral relation previously measured by \cite{Fall1980}, \cite{Romanowsky2012} and OG14. We find this deviation to be consistent with CDM theory once we account for the decrease in $\fM$ with decreasing $\Mb$. This has the effect of bending the $\Mb$-$\jb$ relation at the low $\Mb$ end. Lastly \Hi and stellar $M$-$j$ relations are presented separately, displaying a significantly steeper trend for \Hi, explainable by the change in $\fg$ with $\Mb$.
Plotted as a function of $\Mb$, the \Hi and stellar relations fall roughly parallel with 2.5 times more sAM in the \Hi components.

This work demonstrates the enormous importance of 21cm radio observations of \Hi in measuring the AM of dwarf galaxies. Already in more massive main sequence galaxies, such observations are crucial because most AM resides at large radii that are often \Hi dominated. In dwarf galaxies the situation is even more pronounced since these galaxies are \Hi dominated at virtually all radii. With future radio telescopes such as the Square Kilometre Array (SKA) and its pathfinders coming online, larger, deeper and more highly resolved samples will allow a much more complete analysis of the $M$-$j$ plane across a wide mass range.®




\end{document}